\documentclass{PoS}

\usepackage{amssymb,amsmath,amsbsy}
\usepackage{mathrsfs}
\usepackage{mathtools}
\usepackage{float}

\newcommand{\newc}{\newcommand}
\newc{\be}{\begin{equation}}
\newc{\ee}{\end{equation}}
\newc{\bal}{\begin{align}}
\newc{\eal}{\end{align}}
\newc{\ba}{\begin{eqnarray}}
\newc{\ea}{\end{eqnarray}}
\newc{\bea}{\begin{eqnarray*}}
\newc{\eea}{\end{eqnarray*}}
\newc{\D}{\partial}
\newc{\shh}{\lambda_{h}}
\newc{\shf}{\lambda_{h\phi}}
\newc{\sff}{\lambda_{\phi}}
\newc{\sss}{\lambda_{s}}
\newc{\ssisi}{\lambda_{\sigma}}
\newc{\sssi}{\lambda_{s \sigma}}
\newc{\shs}{\lambda_{h s}}
\newc{\shsi}{\lambda_{h \sigma}}
\newc{\sfs}{\lambda_{\phi s}}
\newc{\sfsi}{\lambda_{\phi \sigma}}
\newc{\shz}{\lambda_{h}^{(0)}}
\newc{\shfz}{\lambda_{h\phi}^{(0)}}
\newc{\sfiz}{\lambda_{\phi}^{(0)}}
\newc{\sssz}{\lambda_{s}^{(0)}}
\newc{\ssfz}{\lambda_{s\phi}^{(0)}}
\newc{\som}{\sin\omega}
\newc{\com}{\cos\omega}
\newc{\sth}{\sin\theta}
\newc{\cth}{\cos\theta}
\newc{\stom}{\sin^2\omega}
\newc{\ctom}{\cos^2\omega}
\newc{\stth}{\sin^2\theta}
\newc{\ctth}{\cos^2\theta}
\newc{\ie}{{\it i.e.} }
\newc{\eg}{{\it e.g.} }
\newc{\etc}{{\it etc.} }
\newc{\etal}{{\it et al.}}

\newcommand{\eV}{\; \mathrm{eV}}

\newcommand{\GeV}{\; \mathrm{GeV}}
\newcommand{\TeV}{\; \mathrm{TeV}}
\newcommand{\lapproxeq}{\lower .7ex\hbox{$\;\stackrel{\textstyle
<}{\sim}\;$}}
\newcommand{\gapproxeq}{\lower .7ex\hbox{$\;\stackrel{\textstyle
>}{\sim}\;$}}
\newcommand{\stackdown}[2]{\lower 1.4ex\hbox{$\;\stackrel{\textstyle{#1}}
{\scriptstyle{#2}}\;$}}

\title{Dark matter and neutrino masses from a classically scale-invariant multi-Higgs
portal}

\ShortTitle{Dark matter and neutrino masses from a classically scale-invariant multi-Higgs
portal}

\author{\speaker{Alexandros Karam}\\
        University of Ioannina, Greece\\
        E-mail: \email{alkaram@cc.uoi.gr}}

\author{Kyriakos Tamvakis\\
        University of Ioannina, Greece\\
        E-mail: \email{tamvakis@uoi.gr}}

\abstract{We present a classically scale-invariant model where the dark matter, neutrino and electroweak mass scales are dynamically generated from dimensionless couplings. The Standard Model gauge sector is extended by a dark $SU(2)_X$ gauge symmetry that is completely broken through a complex scalar doublet via the Coleman-Weinberg mechanism. The three resulting dark vector bosons of equal mass are stable and can play the role of dark matter. We also incorporate right-handed neutrinos which are coupled to a real singlet scalar that communicates with the other scalars through portal interactions. The multi-Higgs sector is analyzed by imposing theoretical and experimental constraints. We compute the dark matter relic abundance and study the possibility of the direct detection of the dark matter candidate from XENON 1T.}

\FullConference{Proceedings of the Corfu Summer Institute 2015 "School and Workshops on Elementary Particle Physics and Gravity"\\
		1-27 September 2015\\
		Corfu, Greece}

\begin{document}

\section{Introduction}

In 2012, the Higgs boson~\cite{Englert1964,Higgs1964,Higgs1964a,Guralnik1964} of the Standard Model (SM) was at last discovered by the ATLAS~\cite{Aad2012} and CMS~\cite{Chatrchyan2012} detectors of the Large Hadron Collider (LHC) at CERN. With its primary goal achieved, LHC could now focus its searches to beyond the Standard Model (BSM) physics. Unfortunately, the first run of LHC ended with no new breakthroughs. Nevertheless, Run II is now underway and is full of promise for unexpected surprises.

Despite its enormous success for over 40 years, the SM is inadequate for explaining some phenomena and it also has a few shortcomings of its own. Some problems of theoretical nature are the \textit{hierarchy problem} and the \textit{vacuum stability problem}. Problems of phenomenological nature are the observed non-zero masses of the SM neutrinos and the existence of dark matter (DM) in the Universe. In this talk we will present possible solutions to these problems.

In the SM, the only dimensionful parameter is the $\mu^2_{SM}$ mass parameter in the Higgs potential, responsible for the electroweak symmetry breaking. Most BSM models posit the existence of large new physics scales $\Lambda_{NP}$, associated with new heavy particles. These particles give huge quadratic corrections ($\Delta M^2_h$) to the renormalized Higgs mass ($M^2_h$). As a result, the bare Higgs mass ($M^2_0$) has to be extremely fine-tuned in order to obtain the measured value $M_h = 125.09 \pm 0.24 \GeV$~\cite{Aad2015}. This is called the hierarchy or naturalness problem. If one sets $\mu_{SM}=0$, then the resulting theory is manifestly classically scale-invariant (CSI) and free from quadratic sensitivity~\cite{Bardeen:1995kv}. Symmetry breaking can then be realized via the Coleman-Weinberg mechanism (CWM)~\cite{Coleman1973}. For the CWM to be successful, though, new bosonic degrees of freedom have to be added to the SM particle content.

The other parameter entering the Higgs potential is the Higgs field self-coupling $\shh$. Studies of its renormalization group evolution have shown~\cite{Buttazzo2013} that it becomes negative above energies of $\mathcal{O}(10^{10}\GeV)$, thus rendering the vacuum metastable. This situation can be remedied if we add scalar fields to the SM and couple them to the Higgs field, since the new portal couplings would contribute positively to the renormalization group equation (RGE) of $\shh$.

Another motivation for adding new degrees of freedom to the SM stems from the strong observational evidence for oscillations between the SM neutrinos, implying non-zero masses and mixings. In a CSI framework one can generate neutrino masses at tree-level by introducing singlet right-handed neutrinos and coupling them to a new singlet scalar that obtains a vacuum expectation value (vev). The product of the right-handed neutrino Yukawa coupling(s) with the singlet vev results in a mass term for the right-handed neutrinos, while the SM neutrinos can obtain their masses through a type-I seesaw mechanism.

Nowadays, dark matter (DM) is believed to constitute nearly $27\%$ of the energy content of the Universe. Its measured relic abundance implies cold DM with annihilation cross section around the electroweak scale. This is the well-known weakly interacting massive particle \textit{(WIMP) miracle}. A DM particle needs to be stable so that it cannot decay to SM particles. This can be achieved by imposing a discrete symmetry under which the DM particles are charged but the SM ones are not. A new discrete symmetry is best motivated as resulting from the breaking of a gauge symmetry.

In this talk, we will present a model~\cite{Karam:2015jta} where we considered the CSI SM, extended by a dark $SU(2)_X$ gauge symmetry~\cite{Hambye2013,Carone2013,Khoze2014,Pelaggi2015}. We also incorporated three right-handed neutrinos and a real scalar singlet in order to implement a type-I seesaw mechanism for the neutrino mass generation. The extra $SU(2)_X$ gauge symmetry gets completely broken by a complex scalar doublet that obtains a vev by means of the CWM. The three new vector bosons obtain equal masses and are stable due to a remnant $Z_2\times Z_2'$ symmetry; therefore they can play the role of WIMP DM. The additional scalar singlet and the Higgs field also get a vev and obtain mass terms through their portal interactions with the dark scalar doublet, in a cascading symmetry breaking effect.

\section{The model}
We start by presenting the model and studying its properties.

\subsection{The tree-level scalar potential}
The scalar sector consists of the Higgs doublet $H$, the dark $SU(2)_X$ doublet $\Phi$ and the real singlet $\sigma$. The most general renormalizable CSI tree-level scalar potential involving these three fields has the form
\be 
  V_0\, =\,\shh ( H^{\dagger} H )^2 + \sff ( \Phi^{\dagger} \Phi )^2 +  \frac{\ssisi}{4} \sigma^4 - \shf ( H^{\dagger} H ) ( \Phi^{\dagger} \Phi )  - \frac{\sfsi}{2} ( \Phi^{\dagger} \Phi ) \sigma^2 
   + \frac{\shsi}{2} ( H^{\dagger} H ) \sigma^2.\label{tlpot}  
\ee
We chose negative signs for the portal couplings $\shf$ and $\sfsi$ so that $H$ and $\sigma$ can obtain mass terms through the vev of $\Phi$.

Apart from the scalar potential, we also introduce the following Yukawa terms for the three right-handed neutrinos:
\be 
-\mathcal{L}_N = Y^{ij}_{\nu} \bar{L}_i \, i\sigma_2 H^{*} N_j + \text{H.c.} + {Y^{ij}_\sigma} \bar{N}^{c}_i N_j \sigma, 
\label{YukawaTerms}
\ee
where $Y^{ij}_\sigma$ denotes the right-handed Majorana neutrino Yukawa matrix which couples the singlet scalar $\sigma$ with the singlet neutrinos and is assumed diagonal and real. $Y^{ij}_{\nu}$ is the Dirac neutrino Yukawa matrix which couples the SM Higgs doublet $H$ with the left-handed lepton doublet $L_i$ and the right-handed neutrinos $N_j$.

In the unitary gauge, the scalar fields obtain the form 
\be 
H = \frac{1}{\sqrt{2}}\left(\begin{array}{c}
0\\
h
\end{array}\right), \quad  \Phi = \frac{1}{\sqrt{2}}\left(\begin{array}{c}
0\\
\phi
\end{array}\right), \quad \sigma = \sigma ,
\label{UNIFIELDS}
\ee
and the tree-level potential is now given by
\be 
V_{0} (h,\phi,\sigma) = \frac{\shh}{4} h^4 + \frac{\sff}{4} \phi^4 + \frac{\ssisi}{4} \sigma^4 - \frac{\shf}{4} h^2 \phi^2  - \frac{\sfsi}{4} \phi^2 \sigma^2  + \frac{\shsi}{4} h^2 \sigma^2.
\label{treelevelpotential}
\ee
The tree-level potential is bounded from below if the following conditions~\cite{Kannike2012} are satisfied for all energies up to the Planck scale:
\be
\lambda_h\,\geq\,0,\,\lambda_{\phi}\,\geq\,0,\,\lambda_{\sigma}\,\geq\, 0, \label{stabilityconditions1}
\ee
\be\frac{\lambda_{h\phi}}{2\sqrt{\lambda_h\lambda_{\phi}}}\,\leq\,1,\,\,\frac{\lambda_{\phi\sigma}}{ 2\sqrt{\lambda_{\phi}\lambda_{\sigma}}}\,\leq\,1\,\,\,\frac{{-\lambda_{h\sigma}}}{ 2\sqrt{\lambda_h\lambda_{\sigma}}}\,\leq\,1, \label{stabilityconditions2}
\ee
\be 
4\lambda_h\lambda_{\phi}\lambda_{\sigma}\,-\left(\lambda_{h\phi}^2\lambda_{\sigma}+ \lambda_{\phi\sigma}^2\lambda_h+\lambda_{h\sigma}^2\lambda_{\phi}\right)\,+ \lambda_{h\phi}{\lambda_{\phi\sigma}}\lambda_{h\sigma}\,\geq\,0. \label{stabilityconditions3}
\ee
Following the Gildener-Weinberg (GW) approach~\cite{Gildener1976}, we may parametrize the scalar fields as 
\be 
h = \varphi N_1, 
\quad 
\phi = \varphi N_2, 
\quad 
\sigma = \varphi N_3,
\ee
where $N_i$ is a unit vector in the three-dimensional field space. Along a particular flat direction $N_i=n_i$ the conditions for an extremum are~\cite{Gildener1976}
\be 
\left.\frac{\partial V_0}{\partial N_i}\right|_{\bf{n}}\,=\,V_0(\mathbf{n})\,=\,0\, {\label{MINI}}
\ee
and result in three tadpole equations and one flatness equation
\ba 
2\shh n_1^2 &=& \shf n_2^2 - \shsi n_3^2, \label{FlatDirectionA} \\
2\sff n_2^2 &=& \shf n_1^2 + \sfsi n_3^2 ,\label{FlatDirectionB} \\
2\ssisi n_3^2 &=& \sfsi n_2^2 - \shsi n_1^2, \label{FlatDirectionC} 
\ea
\be
\shh n_1^4 + \sff n_2^4 + \ssisi n_3^4 - \shf n_1^2 n_2^2 - \sfsi n_2^2 n_3^2 + \shsi n_1^2 n_3^2 = 0, \label{FlatDirectionD}
\ee
with $n_1^2 + n_2^2 + n_3^2 = 1$.

\subsection{The scalar masses}
On the flat direction, the shifted scalar fields can be written as
\be
h\,=\,(\varphi+v)\,n_1, \quad \phi\,=\,(\varphi+v)\,n_2, \quad\sigma\,=\,(\varphi+v)\,n_3 .
\ee
Then, in the $(h, \phi, \sigma)$ basis, the scalar mass matrix at tree level has the form
\be 
\mathcal{M}_0^2 = v^2 
\left( \begin{array}{ccc}
2\shh n_1^2    & -n_1 n_2 \shf  & +n_1 n_3 \shsi  \\
-n_1 n_2 \shf  & 2\sff n_2^2    & -n_2 n_3 \sfsi  \\
+n_1 n_3 \shsi & -n_2 n_3 \sfsi &  2\ssisi n_3^2
\end{array}
\right)\label{massmatrix}
\ee
The above mass matrix can be diagonalized by means of an orthogonal rotation matrix, 
\be 
{\cal{M}}_{d}^2\,=\,{\cal{R}}\,{\cal{M}}_0^2\,{\cal{R}}^{-1},
\ee
where
\be 
{\cal{R}}^{-1}\,=\,\left( \begin{array}{ccc}
\cos\alpha\cos\beta  & \sin\alpha & \cos\alpha\sin\beta  \\
-\cos\beta\cos\gamma\sin\alpha + \sin\beta\sin\gamma      & \cos\alpha\cos\gamma   & -\cos\gamma\sin\alpha\sin\beta - \cos\beta\sin\gamma \\
-\cos\gamma\sin\beta - \cos\beta\sin\alpha\sin\gamma  & \cos\alpha\sin\gamma & \cos\beta\cos\gamma - \sin\alpha\sin\beta\sin\gamma
\end{array}
\right).\label{rotationmatrix}
\ee
If we parametrize the individual vevs with respect to the total vev $v$ according to
\be 
\begin{array}{l} 
v_h = v \sin\alpha = v n_1 , \\
v_\phi = v  \cos\alpha\cos\gamma = v n_2 ,\\
v_\sigma = v \cos\alpha\sin\gamma = v n_3 ,
\end{array}
\label{parametrize}
\ee
then the remaining angle $\beta$ has to be given by the relation
\be
\tan2\beta = \frac{v_h v_\phi v_\sigma v \left( \shsi + \shf \right)}{\left( \sff + \ssisi + \sfsi \right) v_\phi^2 v_\sigma^2 - \shh v_h^2 v^2} \label{angles}
\ee
in order for ${\cal{M}}_d^2$ to be diagonal. For small mixing among the scalars the vevs are hierarchically structured ($v_\phi > v_\sigma > v_h$) and the mass eigenvalues have the form
\ba 
M^2_{h_1} &=& 2\shh v_h^2 \cos^2\alpha \cos^2\beta + \shf v_h v_\phi \cos^2\beta \cos\gamma \sin 2\alpha - \shsi v_h v_\sigma \cos\alpha \cos\gamma \sin 2\beta + \ldots , \\
M^2_{h_2} &=& 0 , \\
M^2_{h_3} &=& 2\ssisi v^2_\sigma \cos^2\beta \cos^2\gamma + \sfsi v_\phi v_\sigma \cos^2\beta \sin 2\gamma + \shsi v_h v_\sigma \cos\alpha \cos\gamma \sin 2\beta + \ldots . 
\ea
One of these masses ($M_{h_2}$) is exactly zero at tree level due to the GW conditions \eqref{MINI} and our particular parametrization of the vevs \eqref{parametrize}. Nevertheless, as we will see later on, this pseudo-Goldstone boson of broken scale invariance which we call ``darkon", will receive large corrections at the one loop level. Finally, we will identify the first eigenvalue with the Higgs mass central value $M_{h_1} = 125.09 \GeV$.

\subsection{Neutrinos}
After symmetry breaking, the Yukawa terms in \eqref{YukawaTerms} relevant for the neutrinos masses are
\be 
\frac{Y^{ij}_{\nu}}{\sqrt{2}} v_h \nu_i \, i\sigma_2  N_j + \text{H.c.} + {Y^{ij}_\sigma}v_{\sigma} \bar{N}^{\;c}_i N_j \,.
\ee
and result in the eigenvalues 
\be 
M_N\,\approx\,{Y_\sigma}v_{\sigma},\quad m_{\nu}\,\approx\,\frac{v_h^2}{4v_{\sigma}}\,\frac{Y^2_\nu}{Y_\sigma},
\ee
where we omitted the indices of the Yukawa matrices. Assuming the right-handed neutrino Yukawa coupling $Y_\sigma$ to be $\mathcal{O}(0.1)$, the vev of the singlet scalar $v_\sigma$ to be $\mathcal{O}(1 \TeV)$ and the left-handed neutrino Yukawa coupling $Y_\nu$ to be $\mathcal{O}(10^{-6} \eV)$ we obtain $M_N \sim \mathcal{O}(100 \GeV)$ and $m_{\nu} \sim \mathcal{O}(0.1 \eV)$.

\subsection{The one-loop potential}
The one-loop potential becomes dominant along the minimum flat direction at a renormalization scale $\Lambda$ where the scalar couplings satisfy the equality in \eqref{stabilityconditions3}. It has the form
\be 
V_1(\mathbf{n}\varphi) = A \varphi^4 + B \varphi^4 \log\frac{\varphi^2}{\Lambda^2},
\ee
where, in the $\overline{MS}$ scheme, the coefficients $A$ and $B$ are given by
\be 
\begin{split}
A &= \frac{1}{64\pi^2 \upsilon^4} \left[ \sum_{i=1,3} M^4_{h_i} \left( -\frac{3}{2} + \log \frac{M^2_{i}}{\upsilon^2} \right) +6 M^4_W \left( -\frac{5}{6} + \log \frac{M^2_W}{\upsilon^2}  \right) + 3 M^4_Z \left( -\frac{5}{6} + \log \frac{M^2_Z}{\upsilon^2} \right) \right.
  \\ & \left. + 9 M^4_X \left( -\frac{5}{6} + \log \frac{M^2_X}{\upsilon^2}  \right) - 12 M^4_t \left( -1 + \log \frac{M^2_t}{\upsilon^2}  \right) - 2 \sum^3_{i=1} M^4_{N_i} \left( -1 + \log \frac{M^2_{N_i}}{\upsilon^2}  \right) \right],
\end{split}
\ee
\be 
B = \frac{1}{64\pi^2 \upsilon^4} \left(\sum_{i=1,3} M^4_{h_i}  +6 M^4_W  + 3 M^4_Z 
  + 9 M^4_X  - 12 M^4_t  - 2 \sum^3_{i=1} M^4_{N_i}  \right).
\ee
Minimizing the one-loop effective potential we obtain
\be 
V_1(\mathbf{n}\varphi) = B \varphi^4 \left[ \log\frac{\varphi^2}{\upsilon^2} - \frac{1}{2} \right].
\label{onelooppotential}
\ee
The one-loop potential is bounded from below if $B>0$. Then, the darkon mass is no longer zero but is given by
\be 
M^2_{h_2} = \frac{1}{8\pi^2 \upsilon^2} \left(M^4_{h_1}+M^4_{h_3}  +6 M^4_W  + 3 M^4_Z 
  + 9 M^4_X  - 12 M^4_t  - 6M^4_{N}  \right). 
\label{PLBMASS}
\ee
\section{Phenomenological analysis}
Next, we will impose theoretical and experimental constraints on the model and examine its phenomenological viability.

\subsection{Theoretical constraints}
Vacuum stability is guaranteed if the conditions \eqref{stabilityconditions1}-\eqref{stabilityconditions3} and $B>0$ are satisfied for all energies up to the Planck scale. The latter is equivalent to 
\be 
M^4_{h_3} + 9 M^4_X - 6 M^4_{N} > \left( 317.26 \,\, \GeV \right)^4 .
\label{massesinequality}
\ee
We check that the stability conditions \eqref{stabilityconditions1}-\eqref{stabilityconditions3} hold up to the Planck scale by numerically solving the full two loop RGEs for all the couplings 
{\allowdisplaybreaks \begin{align} 
\beta_{g_1} & =  
\frac{41}{10} g_{1}^{3} + \frac{1}{(4\pi)^2} \frac{1}{50} g_{1}^{3} \Big( 199 g_{1}^{2} + 135 g_{2}^{2} + 440 g_{3}^{2} -85 y_t^2 \Big) \\ 
\beta_{g_2} & =  -\frac{19}{6} g_{2}^{3} + \frac{1}{(4\pi)^2} \frac{1}{30} g_{2}^{3} \Big( 27 g_{1}^{2} + 175 g_{2}^{2} + 360 g_{3}^{2} -45 y_t^2 \Big)  \\  
\beta_{g_3} & =  -7 g_{3}^{3}  + \frac{1}{(4\pi)^2}\frac{1}{10} g_{3}^{3} \Big( 11 g_{1}^{2} + 45 g_{2}^{2} - 260 g_{3}^{2} - 20 y_t^2  \Big)\\  
\beta_{g_X} & =  -\frac{43}{6} g_{X}^{3}  -\frac{1}{(4\pi)^2}\frac{259}{6} g_{X}^{5} \\
\beta_{y_t} & = y_t \left( \frac{9}{2} y_t^2 -\frac{17}{20} g_{1}^{2} -\frac{9}{4} g_{2}^{2} -  8 g_{3}^{2} \right) \\ 
\beta_{Y_\sigma} & =  
4 \, Y_\sigma \, \mbox{Tr}\Big({Y_\sigma  Y_\sigma^*}\Big)  + 12 \, {Y_\sigma  Y_\sigma^*  Y_\sigma} \\
\beta_{\lambda_h} & =  -6 y_t^4 +24 \lambda_{h}^{2}  +\lambda_h \left( 12 y_t^2 -\frac{9}{5} g_{1}^{2}  -9 g_{2}^{2} \right) +\frac{27}{200} g_{1}^{4} +\frac{9}{20} g_{1}^{2} g_{2}^{2} +\frac{9}{8} g_{2}^{4}  +2 \lambda_{h\phi}^{2} +\frac{1}{2} \lambda_{h\sigma}^{2} \label{lambdahRGE}  \\ 
\beta_{\lambda_\phi} & =  \frac{9}{8} g_{X}^{4} -9 g_{X}^{2} \lambda_\phi +
 24 \lambda_{\phi}^{2}  + 2 \lambda_{h\phi}^{2}   + \frac{1}{2} \lambda_{\phi\sigma}^{2}   \label{lambdaphiRGE}  \\ 
\beta_{\lambda_\sigma} & =  
-64 \mbox{Tr}\Big({Y_\sigma  Y_\sigma^*  Y_\sigma  Y_\sigma^*}\Big) + 16 \lambda_\sigma \mbox{Tr}\Big({Y_\sigma  Y_\sigma^*}\Big)  + 18 \lambda_{\sigma}^{2}  + 2 \lambda_{h\sigma}^{2}  + 2 \lambda_{\phi\sigma}^{2} \label{lambdasigmaRGE}  \\ 
\beta_{\lambda_{h\phi}} & =  \lambda_{h\phi} \left(6 y_t^2 + 12 \lambda_h + 12 \lambda_\phi -4 \lambda_{h\phi}    -\frac{9}{10} g_{1}^{2} -\frac{9}{2} g_{2}^{2} -\frac{9}{2} g_{X}^{2}   \right)
  + \lambda_{h\sigma} \lambda_{\phi\sigma}  \\   
\beta_{\lambda_{\phi\sigma}} & =  \lambda_{\phi\sigma} \left( 8 \mbox{Tr}\Big({Y_\sigma  Y_\sigma^*}\Big) + 12 \lambda_\phi + 6 \lambda_\sigma  -4 \lambda_{\phi\sigma}   -\frac{9}{2} g_{X}^{2} \right)
 +4 \lambda_{h\sigma} \lambda_{h\phi} \\  
\beta_{\lambda_{h\sigma}} & =  \lambda_{h\sigma} \left( 6 y_t^2 +8 \mbox{Tr}\Big({Y_\sigma  Y_\sigma^*}\Big) +12 \lambda_h +6 \lambda_\sigma  + 4 \lambda_{h\sigma} -\frac{9}{10} g_{1}^{2}  -\frac{9}{2} g_{2}^{2}  \right)
    +4 \lambda_{h\phi} \lambda_{\phi\sigma},
\end{align} }
where we defined $\beta_\kappa \equiv \left( 4 \pi \right)^2 \frac{d \kappa}{d \ln \mu}$.

We also require perturbativity of the couplings by demanding that
\be 
\text{all couplings} < 2\,\pi.
\label{perturbativity}
\ee
The RGE of the Higgs self-coupling $\shh$ \eqref{lambdahRGE} contains two new contributions from the portal couplings $\shf$ and $\shsi$ that can help it remain positive up the Planck scale. Furthermore, we can avoid Landau poles for the right-handed neutrino Yukawa coupling and the dark gauge coupling if $Y_\sigma(M_N) \lesssim 0.31$ and $g_X(M_X) \lesssim 2.51$ respectively.

\begin{figure}[H]
\centering
{\includegraphics[width=11.5cm]{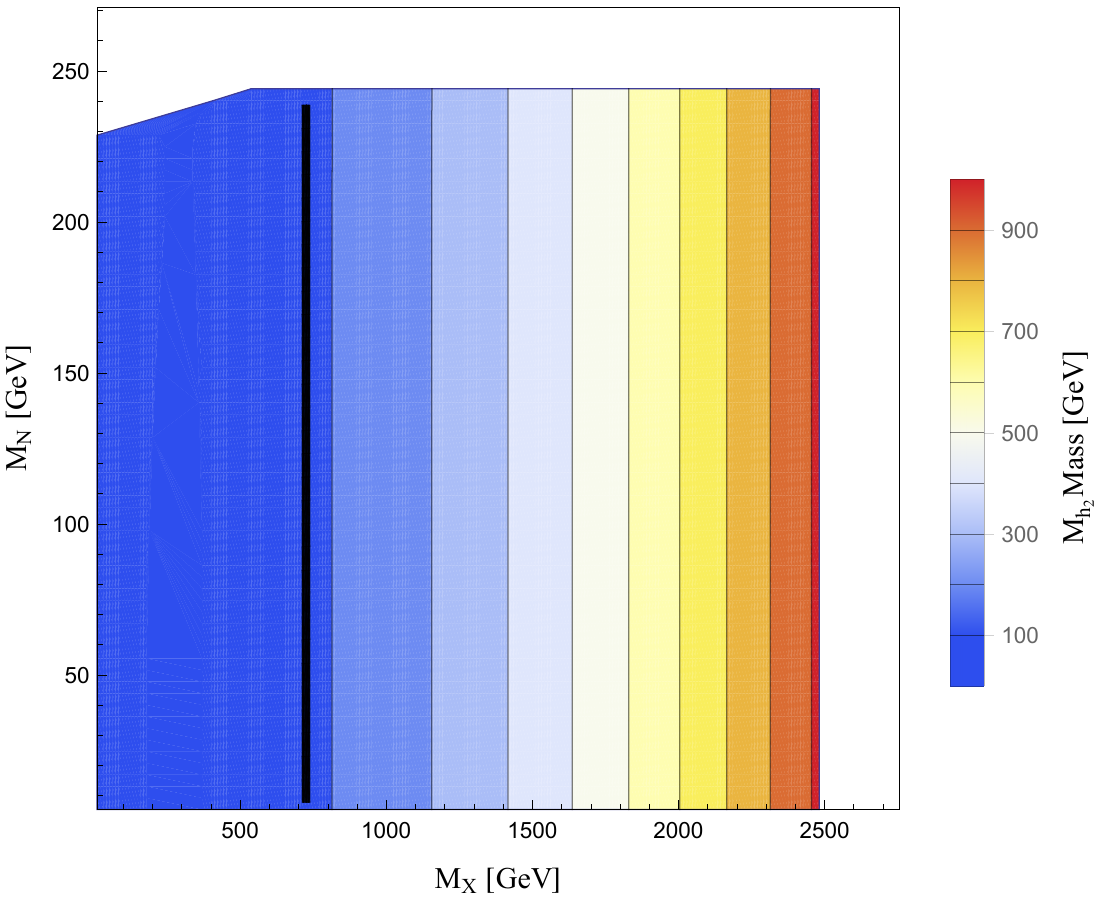}}
\caption{(color online). Scan on the parameter plane $\left( g_X, \, Y_\sigma \right)$ with the rest of the parameters fixed as in Table 1. The color coding indicates the resulting mass of the darkon $M_{h_2}$. The black band corresponds to the points that are able to reproduce the observed DM relic density at $3\sigma$.}
\label{fig:Mh2andRDPlot}
\end{figure}

Since there are many free parameters in this model, we fix most of them as in Table \ref{table:parameters} in order to obtain the measured Higgs mass $M_{h_1}=125.09 \GeV$. This benchmark point satisfies all the stability and perturbativity constraints. Then we scan over the plane $\left( g_X, \, Y_\sigma \right)$ and obtain the mass contours for the darkon mass $M_{h_2}$ shown in Fig. \ref{fig:Mh2andRDPlot}.
\begin{table}[H]
\centering
\begin{tabular}{ | c | c | c | c | c | c | c | c | c | c |}
\hline 
 $v_h[\GeV]$ & $v_\phi[\GeV]$ & $v_\sigma[\GeV]$ & $\shh (\Lambda)$  & $\sff (\Lambda)$ & $\ssisi (\Lambda)$  & $\shf (\Lambda)$ & $\sfsi (\Lambda)$ & $\shsi (\Lambda)$ & $M_{h_3}[\GeV]$ \\ \hline
 $246$ & $2112$ & $770$ & $0.1276$ & $0.004$  & $0.2257$ & $0.0036$ & $0.06$ & $0.001$ & $550.62$ \\ 
\hline  
\end{tabular}
\caption{A benchmark set of values for the parameters involved in the scalar sector that reproduce the measured Higgs mass $M_{h_1}=125.09 \GeV$ and satisfy the stability and perturbativity constraints.}
\label{table:parameters}
\end{table}
\subsection{Experimental constraints}
The three scalar fields of the model obtain vevs and therefore mix. Thus the couplings of the Higgs field to the SM fields get suppressed by the factor $\mathcal{R}_{1 1} = \cos\alpha \cos \beta$. The disparity of the state $h_1$ from the SM Higgs may be readily perceived if we construct the {\textit{signal strength parameter}} $\mu_{h_1}$, which takes the form~\cite{Karam:2015jta}
\be
\mu_{h_1} \simeq \cos^2\alpha\cos^2\beta.
\ee
Using the combined ATLAS~\cite{Aad:2014eva} and CMS~\cite{Khachatryan:2014jba} searches, we can constrain the rotation matrix element
\be 
\mathcal{R}_{11} = \cos\alpha\cos\beta > 0.9.
\label{matrixelementconstraint}
\ee
Due to the vev hierarchy $v_\phi > v_\sigma > v_h$ and the small portal couplings considered in the benchmark point of Table \ref{table:parameters} we obtain
\be 
\mathcal{R}_{11} = 0.994,
\ee
which is in accordance with \eqref{matrixelementconstraint}. Therefore the state $h_1$ and the SM Higgs behave very much alike.

\section{Dark matter analysis}
After the dark $SU(2)_X$ gauge symmetry gets completely broken the three dark vector bosons $X^a$ acquire equal masses $M_X=\frac{1}{2}g_X v_\phi$. Their stability is ensured by a remnant $Z_2 \times Z'_2$ symmetry which can be generalized to a global $SO(3)$ symmetry. We can thus consider them as WIMP DM candidates.

\subsection{Boltzmann equation and relic density}
\begin{figure}[H]
\centering
\includegraphics[width=13.5cm]{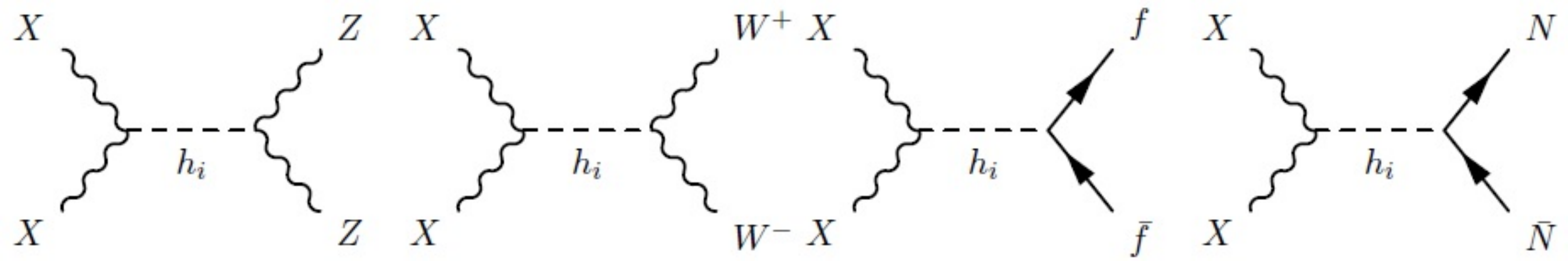}
\caption{DM annihilations to SM particles and right-handed neutrinos.}
\label{fig:DManns1}
\end{figure}

\begin{figure}[H]
\centering
\includegraphics[width=13.5cm]{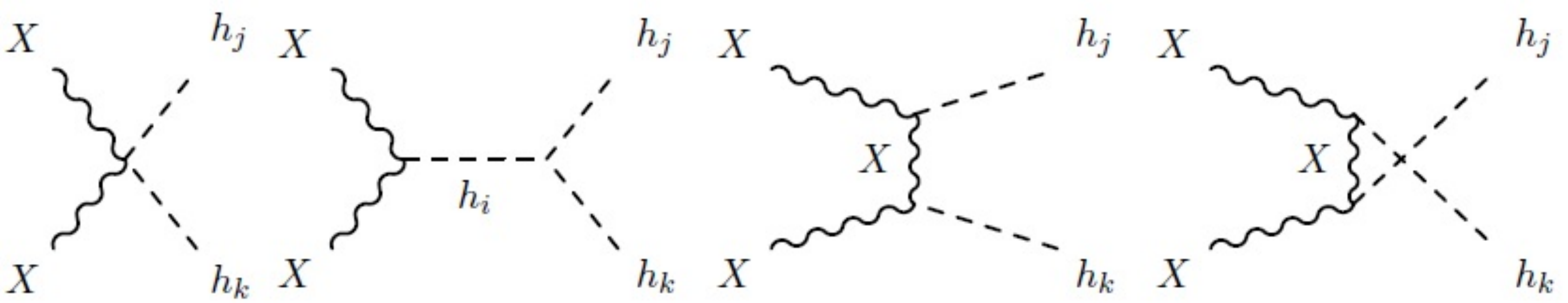}
\caption{DM annihilations to scalars.}
\label{fig:DManns2}
\end{figure}

\begin{figure}[H]
\centering
\includegraphics[width=11.5cm]{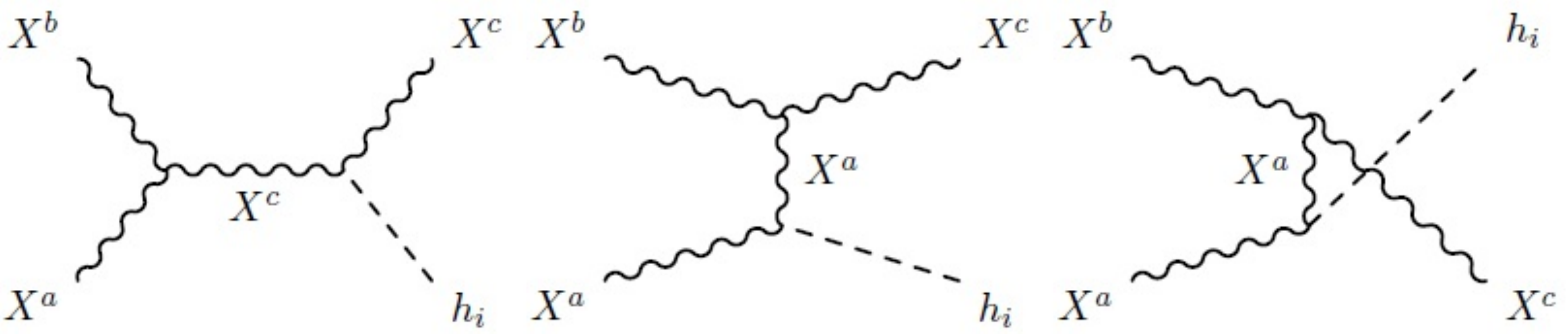}
\caption{DM semiannihilations.}
\label{fig:DMsemi}
\end{figure}

The novel thing about $SU(2)_X$ vector DM is that apart from annihilations, the DM particles can also semiannihilate~\cite{DEramo2010} (see Figs. \ref{fig:DManns1}-\ref{fig:DMsemi} for the relevant Feynman diagrams).
\vspace{0.1cm}
\vspace{0.1cm}
In order to calculate the DM relic abundance we need to solve the Boltzmann equation describing the evolution of the number density $n$ of the DM particles species,
\be 
\frac{d n}{d t} + 3\,H\,n = - \, \frac{\left< \sigma v \right>_{a}}{3}  \left( n^2 - n^2_{eq} \right) - \frac{2\left< \sigma v \right>_{s}}{3} \, n \left( n - n_{eq} \right), \label{BoltzmannEquation}
\ee
where $\left< \sigma v \right>_{a}$ ($\left< \sigma v \right>_{s}$) is the thermally averaged annihilation (semiannihilation) cross section of the DM particles times their relative velocity, $H$ is the Hubble expansion parameter and $n_{eq}$ is the equilibrium number density. It is useful to express \eqref{BoltzmannEquation} in terms of the comoving volume $Y=n/\mathbf{s}$, with $\mathbf{s}$ being the entropy density, as 
\be 
\frac{d Y}{d x} = - \frac{Z_a}{3 x^2} \left( Y^2 - Y^2_{eq} \right) - \frac{2 Z_s}{3 x^2} \left( Y^2 - Y\, Y_{eq} \right), \qquad Z_{a,s} \equiv \frac{\mathbf{s}\left( x=1 \right)}{H \left( x=1 \right)} \left< \sigma v \right>_{a,s} , \label{BoltzmannEquation2}
\ee 
where $ H = \sqrt{\frac{4 \pi^3 g_*}{45}} \, \frac{M^2_X}{M_{\text{P}}}$, $
\mathbf{s} = \frac{2 \pi^2 g_*}{45} \, \frac{M^3_X}{x^3}$, $x=M_X/T$ and $g_*=86.25$ is the relativistic degrees of freedom during the freeze-out $\left( x=x_f \right)$. We find the freeze-out point to have values between $x_f \approx 25-26$. Finally, solving \eqref{BoltzmannEquation2} we obtain the relic density of the vector boson particles 
\be 
\Omega_{X} h^2 = 3\times \frac{1.07\times 10^9 \, \GeV^{-1}}{\sqrt{g_*} \, M_P \, J(x_f)}, \qquad J(x_f) = \int_{x_f}^\infty dx \, \frac{\left< \sigma v \right>_{a} + 2 \left< \sigma v \right>_{s}}{x^2}. \label{RelicDensity}
\ee
In Fig. \ref{fig:Mh2andRDPlot}, the black band corresponds to the DM masses that saturate the observed DM relic density at $ 3\sigma $. For the benchmark point in Table \ref{table:parameters} we find DM masses between $M_X \sim 710-740 \GeV$.

\subsection{Dark matter direct detection}
\begin{figure}[H]
\centering
\includegraphics[width=4.5cm]{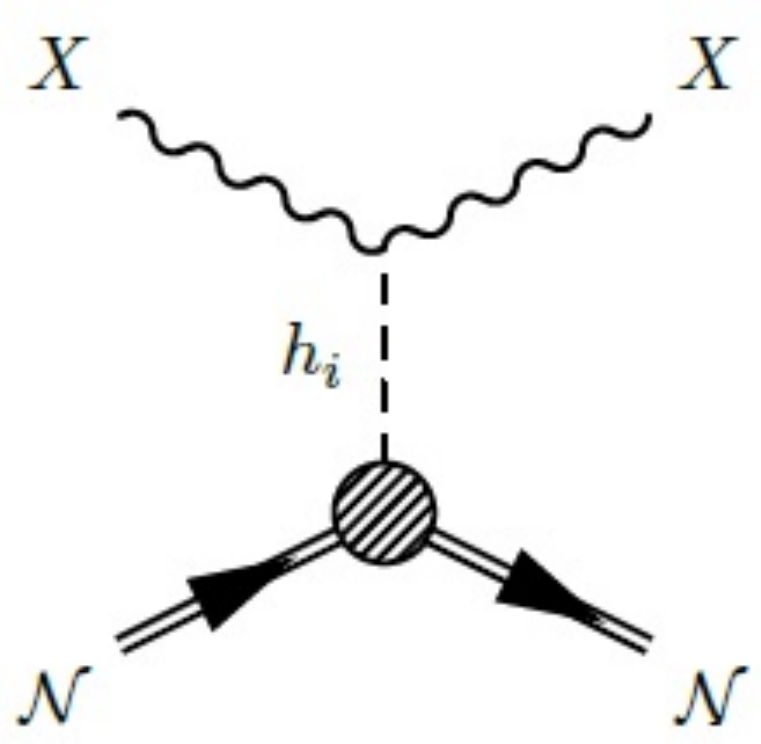}
\caption{Feynman diagram for DM-nucleon elastic scattering.}
\label{fig:DTFeynman}
\end{figure}

The dark vector bosons $X^a$ can in principle interact with nucleons by exchanging scalar bosons $h_i$ (c.f. Fig. \ref{fig:DTFeynman}).
The cross section for this process has the form 
\be 
\sigma_{SI} = \frac{\mu_{\text{red}}^2}{\pi v_h^2 v_\phi^2} \, \left| f_{\mathcal{N}} M_X m_{\mathcal{N}}  \sum_i \frac{\mathcal{R}_{i2} \mathcal{R}_{1i}}{M^2_{h_i}} \right|^2, \label{DTXS}
\ee
where $\mu_{\text{red}} = M_X m_{\mathcal{N}}/\left( M_X + m_{\mathcal{N}} \right)$ is the reduced mass of the DM particles and the nucleons, $m_{\mathcal{N}}=0.939 \GeV $ is the average nucleon mass and $f_{\mathcal{N}}=0.303$ is the nucleon form factor. 

\begin{figure}[H]
\centering
\includegraphics[width=11.5cm]{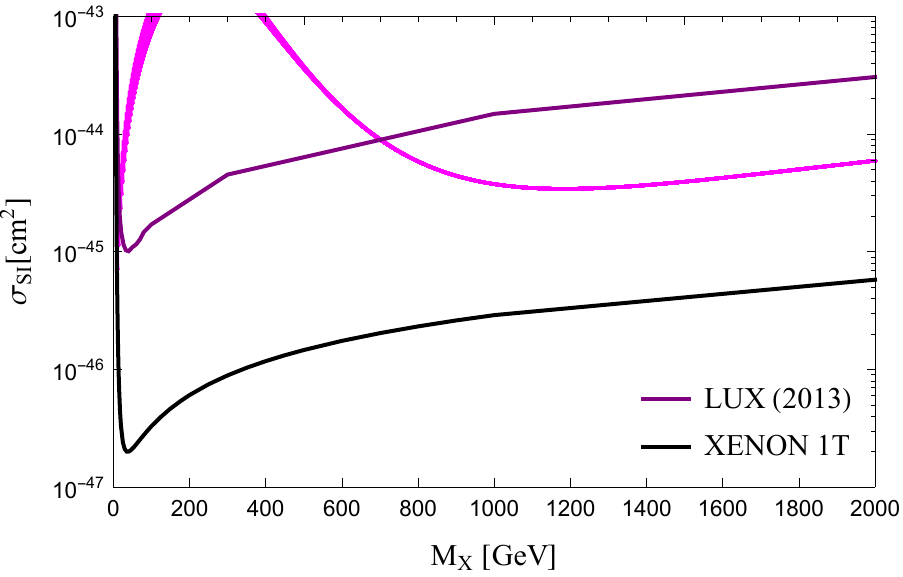}
\caption{(color online). We show the DM-nucleon spin-independent cross section as a function of the dark vector boson mass (magenta band) with varying $M_N$ masses and the rest of the parameters fixed as in Table 1. The purple line signifies the limits set by LUX (2013) while the black line corresponds to the expected limits of XENON 1T.}
\label{fig:DTXS}
\end{figure}

In Fig. \ref{fig:DTXS} we plot the DM masses $M_X$ versus the spin-independent cross section in \eqref{DTXS}. We find that dark vector boson masses above circa $700 \GeV$ evade the bounds set by LUX (2013) but can nevertheless be detected by the currently operating XENON 1T experiment.

\section{Summary and conclusions}
In conclusion, we constructed a classically scale-invariant extension of the Standard Model where the dark matter, neutrino and electroweak scales were dynamically generated and the hierarchy problem was naturally solved. We incorporated three singlet right-handed neutrinos and a singlet scalar field in order to implement a type-I seesaw mechanism. The dark matter consisted of three vector bosons of an extra $SU(2)_X$ gauge symmetry which was broken by a doublet scalar field by means of the Coleman-Weinberg mechanism. Then, through the portal interactions in the scalar potential, a mass scale was communicated to the neutrino sector and the electroweak sector. We examined the tree-level and one-loop scalar potential and saw that the vacuum could be easily stabilized. Employing the Gildener-Weinberg formalism we obtained three massive scalar bosons, one of which was identified with the recently discovered Higgs boson. We proceeded by imposing theoretical and experimental constraints on the parameters of the model and saw that the scalar state identified with the Higgs boson differs only by a very small common suppression factor from it. Finally, we calculated the dark matter relic abundance and the dark matter elastic scattering off a nucleon cross section. Then, using experimental bounds we concluded that the dark matter masses have to lie in the$\TeV$ range.

\acknowledgments
This research has been cofinanced by the European Union (European Social Fund - ESF)
and Greek national funds through the Operational Program Education and Lifelong
Learning of the National Strategic Reference Framework (NSRF) - Research Funding
Program: \textit{ARISTEIA - Investing in the society of knowledge through the European Social Fund.} K.T. would like to thank I. Antoniadis and K. Papadodimas for discussions and hospitality at the CERN Theory Division. A.K. would like to thank Gunnar Ro and Dimitrios Karamitros for useful discussions and correspondence and also the organizers of the ``Summer School and Workshop on the Standard Model and Beyond", Corfu 2015, for the hospitality during his stay and for giving him the opportunity to present this work.

\end{document}